\documentstyle[prl,aps,epsf,floats]{revtex}
\begin{document}
\draft
\twocolumn[\hsize\textwidth\columnwidth\hsize\csname @twocolumnfalse\endcsname
\title{Induced local spin-singlet amplitude and pseudogap   
       in high $T_{c}$ cuprates}
\author{Bumsoo Kyung}
\address{D\'{e}partement de physique and Centre de recherche 
sur les propri\'{e}t\'{e}s \'{e}lectroniques \\
de mat\'{e}riaux avanc\'{e}s. 
Universit\'{e} de Sherbrooke, Sherbrooke, Qu\'{e}bec, Canada J1K 2R1}
\date{January 5, 2001}
\maketitle
\begin{abstract}

    In this paper we show that
local spin-singlet amplitude with $d$-wave symmetry,
$\langle  |\Delta_{d}(0)|^{2} \rangle$, can be induced
by short-range spin correlations
even in the absence of pairing interactions.
Fluctuation theory is formulated to make connection
between pseudogap temperature $T^{*}$, pseudogap size $\Delta_{pg}$
and $\langle  |\Delta_{d}(0)|^{2} \rangle$.
In the present scenario for the pseudogap,
the normal state pseudogap is caused by the induced local
spin-singlet amplitude due to short-range spin correlations,
which compete in the low energy sector with superconducting
correlations to make $T_{c}$ go to zero near half-filling.
Calculated $T^{*}$ falls from a high value onto the $T_{c}$ line and
closely follows mean-field N\'{e}el temperature $T_{N}^{MF}$.
The calculated $\Delta_{pg}$ is
in good agreement with experimental results.
We propose an experiment in which the present scenario can be critically tested.
\end{abstract}
\vskip2pc]
\narrowtext

   The recent discovery of pseudogap in underdoped high  
$T_{c}$ cuprates has challenged condensed matter physicists for several  
years. 
The pseudogap behavior\cite{Timusk:1999} is observed as strong suppression 
of low frequency spectral weight below some characteristic temperature
$T^{*}$ higher than transition temperature $T_{c}$.
This anomalous phenomenon has been observed in 
angle resolved photoemission spectroscopy
(ARPES),\cite{Ding:1996,Loeser:1996}
specific heat,\cite{Loram:1993}
tunneling,\cite{Renner:1998}
NMR,\cite{Takigawa:1991} and
optical conductivity.\cite{Homes:1993}
One of the most puzzling questions in pseudogap phenomena is why 
$T^{*}$ has a completely different doping dependence from $T_{c}$,
in spite of their possibly close relation.

   Among several theoretical proposals\cite{Timusk:1999} to understand the
pseudogap phenomena,
the superconducting (SC) fluctuation scenario\cite{Loktev:2000}  
of pseudogap has received much attention 
from the physics community.
This is because some experiments such as
ARPES and tunneling experiments show that
the normal state pseudogap has the same angular dependence
and magnitude as the SC
gap and that
often the only difference between
the spectra in the pseudogap state and the SC state
is in their linewidths.
The basic idea of this scenario is that SC gap amplitude forms 
at $T_{c}^{MF}$ while its phase coherence is established at $T_{c}$
lower than $T_{c}^{MF}$.
Hence $T^{*} \sim T_{c}^{MF}$ and below this temperature  
SC fluctuations 
become stronger until they diverge at $T_{c}$.

\begin{figure}
 \vbox to 6.5cm {\vss\hbox to -5.0cm
 {\hss\
       {\includegraphics{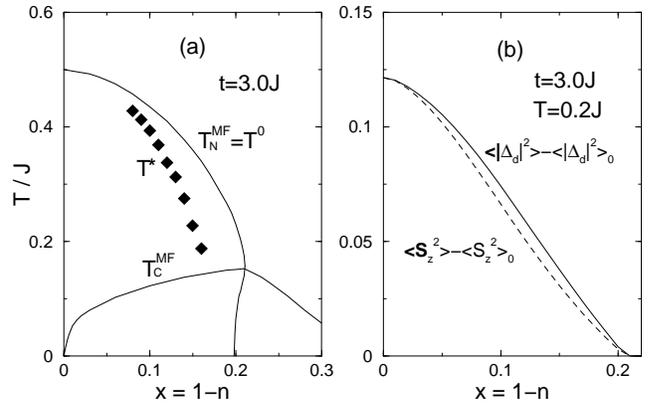}
       }
  \hss}
 }
\caption{(a) Calculated mean-field phase diagram in doping ($x=1-n$)
         and temperature ($T$) plane in the $t-J$ model for $t/J=3.0$.
         $T^{MF}_{N}$ and $T^{MF}_{c}$ are mean-field AF and SC
         ordering temperatures, respectively.
         The filled diamonds are the pseudogap
         temperature determined from the
         single particle spectral function.
         (b) Interaction induced local spin-singlet (solid curve) and spin
         (dashed curve) amplitudes for $t/J=3.0$ and $T=0.2J$.}
\label{fig1}
\end{figure}

   In spite of its success in explaining some features of 
the pseudogap, it suffers from at least three important drawbacks which were
often overlooked in the past.  
First, just below $T^{*}$ there is no experimental evidence 
of characteristic features associated with SC fluctuations
such as  
fluctuating diamagnetic (Meissner) effect,
fluctuating superfluid density and so on.
It appears that $T^{*}$ has nothing to do with superfluid ``rigidity''.
One experiment to strongly support this argument was recently carried  
out by Orenstein's group\cite{Corson:1999}.
In their high-frequency conductivity measurements tracking the 
phase-coherence time $\tau$ in the normal state, the temperature 
$T_{\Theta}^{0}$ where the phase-stiffness of superfluidity disappears
is at most 25 K above $T_{c}$ for underdoped cuprates.  
Second, when SC correlations are treated
on equal footing with antiferromagnetic (AF)
correlations, which is more realistic from both theoretical and
experimental points of view, $T_{c}^{MF}$ goes down to zero
with decreasing doping due to the competition with the AF correlations,
as shown in Fig.~\ref{fig1}(a).
Then the above scenario ($T^{*} \sim T_{c}^{MF}$) is inconsistent  
even with the doping 
dependence of $T^{*}$, which increases 
with decreasing doping.
Apparently experimentally observed $T^{*}$ stays in between $T_{c}^{MF}$
and $T_{N}^{MF}$.
Furthermore, in this situation the origin of the pseudogap itself 
is questionable, because the  
pseudogap appears to be caused by AF fluctuations!
Third, in their recent paper Tallon and Loram\cite{Tallon:2001}
argued, based on experimental results, that 
$T^{*}$ falls from
a high value onto the $T_{c}$ line instead of 
smoothly merging with $T_{c}$ in the slightly overdoped region.
The above scenario for the pseudogap predicts the latter behavior of 
$T^{*}$.
In this paper we demonstrate that
induced local spin-singlet amplitude due to short-range spin correlations
causes a normal state pseudogap with $d$-wave symmetry
even in the absence of pairing interactions.

   First of all we argue that there are 
two energy scales in the problem, 
because the pseudogap appears as a crossover phenomenon according to
experiments.
The low energy (or long distance) physics of AF and SC correlations
is well captured
by a {\it static} mean-field approach, while the relatively high
energy (or short distance) physics of the pseudogap is invisible
in such a study.
Thus we resort to fluctuation theory in order to describe
the dynamical nature of the pseudogap, and to determine
$T^{*}$ and $\Delta_{pg}$.
Note that the present formulation below is different from standard
fluctuation theory based on the (static) mean-field state.
The mean-field result of the $t-J$ model
will be used below solely
to find the onset of
leading correlations\cite{Comment5}, and to compute mean-field AF and SC  
order parameters for the calculation of local spin and spin-singlet
amplitudes.
The mean-field $t-J$ Hamiltonian reads
\begin{eqnarray}
 H_{MF} &=& \sum_{\vec{k}, \sigma}\varepsilon(\vec{k})
                  c^{\dag}_{\vec{k},\sigma}
                  c_{\vec{k},\sigma}
                                             \nonumber  \\
      & &  -2Jm\sum_{\vec{k}}(c^{\dag}_{\vec{k}+\vec{Q},\uparrow}
                              c    _{\vec{k},\uparrow}
                             -c^{\dag}_{\vec{k}+\vec{Q},\downarrow}
                              c    _{\vec{k},\downarrow})
                                             \nonumber  \\
        & &  -Js\sum_{\vec{k}}\phi_{d}(\vec{k})(
              c^{\dag}_{\vec{k},\uparrow}  c^{\dag}_{-\vec{k},\downarrow}
             +c    _{\vec{k},\downarrow}c_{-\vec{k},\uparrow}) \; ,
                                                           \label{eq10}
\end{eqnarray}
%
%
where
$\varepsilon(\vec{k}) \simeq -2tx(\cos k_{x}+\cos k_{y})-\mu$
with $x$ the hole density.
$m$ and $s$ are mean-field AF and SC order parameters determined from
\begin{eqnarray}
 m &=& 
   1/(2N)\sum_{\vec{k},\sigma}\sigma \langle c^{\dag}_{\vec{k}+\vec{Q},\sigma}
                                      c_{\vec{k},\sigma} \rangle \; , 
                                             \nonumber  \\
 s &=&  
    1/N\sum_{\vec{k}}\phi_{d}(\vec{k})
      \langle c_{\vec{k},\uparrow} c_{-\vec{k},\downarrow} \rangle \; ,
                                                           \label{eq11}
\end{eqnarray}
where $N$ is the total number of lattice sites.
$\phi_{d}(\vec{k}) = \cos k_{x}-\cos k_{y}$ and 
$\vec{Q}$ is the (commensurate) AF wave vector $(\pi,\pi)$ in two dimensions.
In this paper we restrict ourselves to a uniform solution which is just enough
for our purpose.
In a mean-field approximation, mean-field order
already sets in when
the correlation length reaches roughly one
lattice spacing.
This forces the above mean-field phase line (Fig.~\ref{fig1}(a))
to be interpreted as
the onset
of the corresponding short-range correlations.
We identify $T_{N}^{MF}$ with another crossover temperature
$T^{0}$ at which some magnetic experiments such as Knight shift 
show their maximum.
For the parameter ($t/J=3.0$) used in this paper, short-range spin  
correlations
disappear at $x=x_{c} \simeq 0.19-0.20$ at low temperature.

   We introduce AF and spin-singlet\cite{Comment15} correlation functions
\begin{eqnarray}
\chi_{sp}(i,\tau) & = & \langle T_{\tau} \sigma_{z}(i,\tau)
                                     \sigma_{z}(0,0) \rangle \; ,
                                             \nonumber  \\
\chi_{pp}(i,\tau) & = & \langle T_{\tau} \Delta_{d}(i,\tau)
                                     \Delta^{\dag}_{d}(0,0) \rangle  \; ,
                                                           \label{eq20}
\end{eqnarray}
where 
$ \sigma_{z}(i) \equiv (c^{\dag}_{i,\uparrow}c_{i,\uparrow}
 -c^{\dag}_{i,\downarrow}c_{i,\downarrow})=2S_{z}(i) $ and 
$ \Delta_{d}(i) = \frac{1}{2}\sum_{\delta}g(\delta)
        (c_{i+\delta,\uparrow}        c_{i,\downarrow}
        -c_{i+\delta,\downarrow}      c_{i,\uparrow}) $ with 
$g(\delta) =  1/2$ for $\delta=(\pm 1,0)$, 
             -1/2  for $\delta=(0,\pm 1)$, and 0 otherwise.  
The AF and spin-singlet correlation functions are related 
to the {\em local} spin and spin-singlet
amplitudes through the sum rules
\begin{eqnarray}
   \frac{T}{N}\sum_{q}\chi_{sp}(q)e^{-i\nu_{m}0^{-}}
&=&  \langle  [\sigma_{z}(0)]^{2} \rangle 
 =   n-2 \langle n_{\uparrow}n_{\downarrow} \rangle  \; ,
                                             \nonumber  \\
   \frac{T}{N}\sum_{q}\chi_{pp}(q)e^{-i\nu_{m}0^{-}}
&=&  \langle  |\Delta_{d}(0)|^{2} \rangle
                                               \; ,
                                                           \label{eq40}
\end{eqnarray}
where $q=(\vec{q},i\nu_{m})$ and $\nu_{m}$ is bosonic Matsubara
frequencies.
$T$ is absolute temperature. 
These sum rules can be easily obtained by taking $\tau=0^{-}$ limit
and setting $i$ to the origin in Eq.~(\ref{eq20}).
In terms of renormalized vertices $U_{sp}$ and $V_{pp}$,\cite{Vilk:1997}
we approximate the AF and spin-singlet correlation functions  
\begin{eqnarray}
\chi_{sp}(q) &=& \frac{2\chi^{0}_{ph}(q)}{1-U_{sp}\chi^{0}_{ph}(q)} \; ,
                                             \nonumber  \\
\chi_{pp}(q) &=& \frac{ \chi^{0}_{pp}(q)}{1-V_{pp}\chi^{0}_{pp}(q)} \; ,
                                                           \label{eq50}
\end{eqnarray}
where the irreducible susceptibilities are defined as
\begin{eqnarray}
\chi^{0}_{ph}(q) &=&-\frac{T}{N}\sum_{k} G^{0}(k-q)G^{0}(k),
                                             \nonumber  \\
\chi^{0}_{pp}(q) &=& \frac{T}{4N}\sum_{k}
         (\phi_{d}(\vec{k})+\phi_{d}(\vec{q}-\vec{k}))^{2}
                                       G^{0}(q-k)G^{0}(k)
                                               \; .
                                                           \label{eq51}
\end{eqnarray}
%
$G^{0}(k)$ is the noninteracting Green's function obtained 
from Eq.~(\ref{eq10}) with $J=0$.
Now two unknown vertices, $U_{sp}$ and $V_{pp}$, are determined 
by the sum rules Eq.~(\ref{eq40}).\cite{Vilk:1997}
Hence, an increase in   
the local spin or spin-singlet amplitude evaluated 
in the interacting state over that in the noninteracting state 
leads to a nonvanishing positive value of $U_{sp}$ or $V_{pp}$, namely,
the enhancement of the corresponding correlation function.
This (non-perturbative sum rule) approach has been shown to be quite 
reliable\cite{Vilk:1997}
as long as short range correlations are concerned. 
In our calculations,                                    
the pseudogap appears when the spin-singlet
correlation length reaches about 1 lattice constant.
Since this method is expected to be accurate up to the intermediate coupling 
regime, it forces the effective bandwidth ($W=8tx$) to be larger 
than the effective interaction strength ($2J$), which leads to $x \ge 0.08$.
In order to determine whether the pseudogap is caused by the spin-singlet
or AF spin fluctuation channel, we separately consider the self-energy
coming from each channel
\begin{eqnarray}
\Sigma_{sp}(k) &=&  UU_{sp}\frac{T}{N}\sum_{q}
                    \chi_{sp}(q)G^{0}(k-q) \; ,
                                             \nonumber  \\
\Sigma_{pp}(k) &=& - \frac{1}{4}VV_{pp}
                \frac{T}{N}\sum_{q}
                                             \nonumber  \\
          & &  (\phi_{d}(\vec{k})+\phi_{d}(\vec{q}-\vec{k}))^{2}
                \chi_{pp}(q)G^{0}(q-k)
                                               \; ,
                                                           \label{eq70}
\end{eqnarray}
where $U=2J$ and $V=J$ from Eq.~(\ref{eq10}).

   First let us begin by showing the interaction-induced local  
spin (dashed curve) and 
spin-singlet (solid curve) amplitudes (Fig.\ref{fig1}(b)) evaluated   
in the mean-field state of the $t-J$ Hamiltonian 
in a region where $s=0$ (or $T > T^{MF}_{c}$). 
Since $s=0$, the 
spin-singlet amplitude is entirely caused by short-range spin correlations
in the absence of pairing interactions.
Although in general a mean-field state is not accurate for  
strongly correlated electron systems,  
certain local and short-range {\em static} quantities such as
double occupancy and
nearest neighbor correlations
are reasonably well captured by the mean-field state particularly with AF order 
(See Ref.~\cite{Kyung:2000-3} for more details). 
In fact the interaction-induced local spin and 
spin-singlet amplitudes (Eq.~(\ref{eq40}))
are determined most crucially by these quantities.
The local spin amplitude
starts to appear when short-range spin correlations begin to develop
and keeps growing with decreasing doping, as can be easily expected
from the mean-field phase diagram (Fig.\ref{fig1}(a)) itself.
Quite unexpectedly, however, 
the local spin-singlet amplitude also increases with decreasing doping  
despite the fact that the mean-field SC order $s$ is absent. 
The increase of local spin-singlet amplitude traces back to  
the growing short-range spin correlations with decreasing doping. 
This same feature was recently studied by the
author\cite{Kyung:2000-3} 
in the context of the Hubbard model.

\begin{figure}
 \vbox to 6.0cm {\vss\hbox to -5.0cm
 {\hss\
       {\includegraphics{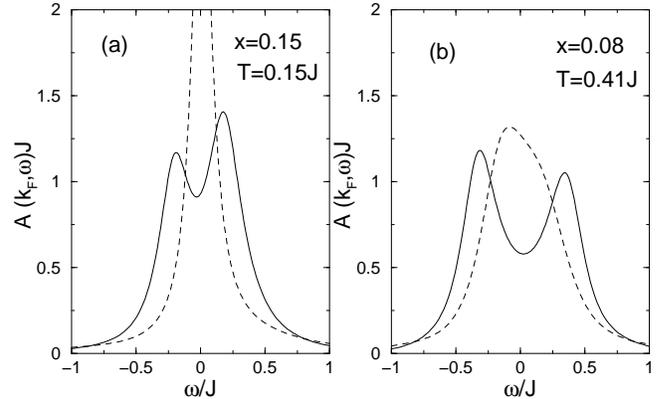}
       }
  \hss}
 }
\caption{Single particle spectral functions from spin-singlet (solid curve) 
         and AF spin fluctuation (dashed curve) channels for 
         (a) $x=0.15$, $T=0.15J$ and for
         (b) $x=0.08$, $T=0.41J$.}
\label{fig2}
\end{figure}

    In Fig.~\ref{fig2} we show the spectral functions 
at $\vec{k}=\vec{k}_{F}$ along $(0,0)-(0,\pi)$ direction
from spin-singlet (solid curve) and AF spin fluctuation (dashed curve) channels
for two doing levels $x=0.15$ and $x=0.08$.
For both densities
the pseudogap appears first in the spin-singlet channel.
This is verified even close to half-filling\cite{Comment30} by 
considering the Hubbard model ($U=8t$)%
\cite{Kyung:2000-3},  
which is not shown in this paper.
The reason why the pseudogap is always caused by the spin-singlet channel is 
that as the local spin amplitude increases with decreasing doping,
the local spin-singlet amplitude also increases at the same time.
Hence the feature found away from half-filling persists down to half-filling.
In our calculations the pseudogap appears when the characteristic
low frequency scale of the spin-singlet correlations is smaller
than temperature (renormalized classical regime).\cite{Vilk:1997}
The pseudogap due to AF spin fluctuations starts to appear
for $x \le 0.10$ with $T \ll T^{*}$.

   Based on the above results we show the pseudogap temperature 
$T^{*}$ (filled diamonds) as a function of doping in Fig.~\ref{fig1}(a).
%
%
$T^{*}$ falls from
a high value onto the $T_{c}$ ($\le T_{c}^{MF}$) line  
instead of sharing a common 
line with $T_{c}$ in overdoped region.
When superconductivity is suppressed by setting $s=0$,
$T^{*}$ vanishes near $x_{c}$ where short-range spin correlations disappear.
It is not surprising to find that $T^{*}$
closely follows $T_{N}^{MF}=T^{0}$, because in our study the pseudogap is 
caused by induced local spin-singlet amplitude  
due to short-range spin correlations, 
which is reasonably well captured by the mean-field state with AF order.
All these features are at least qualitatively consistent with the 
findings by Tallon and Loram.\cite{Tallon:2001}
Although quantitative agreement with the Hubbard and $t-J$ models
is achieved only for very strong coupling 
($U \gg  t$ or $J \ll t$),\cite{Jeckelmann:1998}
it is instructive to compare the calculated $T^{*}$ with the recent calculations
obtained from the {\it dynamical} cluster approximation (DCA) for the Hubbard 
model by Jarrell {\em et al.}\cite{Jarrell:2000}.
Their $T^{*}$ near half-filling is about $0.09t$ for $U=8t$, while ours
is $0.15-0.16t$ for $U=4t^{2}/J=12t$.  
The reasonable agreement with the more systematic approach is 
encouraging in light of the drastically simple approximation used 
in the paper, namely, replacing the strongly correlated hopping 
term of the $t-J$ model by $tx$ and using the mean-field state 
to compute local correlations.

\begin{figure}
 \vbox to 6.5cm {\vss\hbox to -5.0cm
 {\hss\
       {\includegraphics{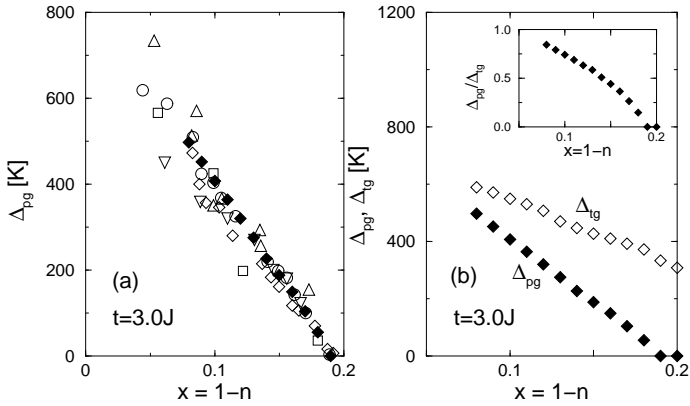}
       }
  \hss}
 }
\caption{(a) Calculated pseudogap size at $T=0$
         denoted as filled diamonds for $t/J=3.0$ and $J=125$ meV.
         The open circles, diamonds, squares, up-triangles, and down-triangles
         are the pseudogap size extracted from susceptibility,
         heat capacity, ARPES, NMR, and resistivity measurements, respectively,
         by Tallon and Loram.\protect\cite{Tallon:2001}
         (b) Total excitation gap $\Delta_{tg}$ (open diamonds) and 
         pseudogap $\Delta_{pg}$ (filled diamonds) 
         at $T=0$. The inset shows their relative ratio.}
\label{fig3}
\end{figure}

  Figure~\ref{fig3}(a) shows the pseudogap size $\Delta_{pg}$  
(filled diamonds) which 
is defined as half of the peak-to-peak distance in Fig.~\ref{fig2}
by setting $s=0$ at $T=0$.
In the same figure the pseudogap energy extracted
from various experiments by Tallon and Loram\cite{Tallon:2001}
is also shown as empty symbols for comparison . 
$\Delta_{pg}$ vanishes near $x_{c}$,
suggesting the presence of a quantum critical point
at a critical doping.
The agreement between our results and experiments appears remarkable 
for such a simple approximation.
The linear vanishing of $\Delta_{pg}$ near $x_{c}$ is closely
related to the corresponding behavior of the induced local
spin-singlet amplitude.

   The total excitation gap (or ARPES leading edge gap or SC gap)
in the SC state, $\Delta_{tg}$,
can be also calculated.
$\Delta_{tg}$ at $T=0$ is always larger than $\Delta_{pg}$ at $T=0$
due to the additional
contribution to the local spin-singlet amplitude from $s \ne 0$,
which is shown
in Fig.~\ref{fig3}(b).
$\Delta_{pg}$, $\Delta_{tg}$ and
their relative ratio $\Delta_{pg} / \Delta_{tg}$ should
be all monotonically decreasing functions of doping,
as shown in the inset of Fig.~\ref{fig3}(b).
Since the SC order parameter vanishes at $T_{c}$
(at $T^{MF}_{c}$ in this paper), the SC gap below $T_{c}$ continuously
evolves into the normal state pseudogap above $T_{c}$ with
the same momentum dependence and magnitude.

   In order to confirm the present scenario for the pseudogap,
we propose an experiment in which
long-range superconductivity is completely destroyed by a phase sensitive
external perturbation such as a strong magnetic field.
Or this can be also done in a vortex core with a relatively weak magnetic
field.
Then our scenario predicts that
the underlying ground state will manifest itself as an insulator
with a pseudogap or spin gap.
The normal state pseudogap $\Delta_{pg}$
should be observable down to $T=0$ for $x \le x_{c}$
vanishing near $x_{c}$, and its size is given by
Fig.~\ref{fig3}(a).
The present scenario for the pseudogap predicts that a normal state pseudogap
is likely to appear when short-range spin correlations are well
established and are not masked by long-range (AF or SC) order.

   The pseudogap appears here only as the suppression of    
low frequency spectral weight   
in certain physical quantities 
which are obtained through $A(\vec{k},\omega)$ or its convolution 
with a relevant vertex.
It does not appear as a thermodynamic phase with broken symmetry.
The pseudogap obtained in this paper is different  
from the spinon gap found 
in the previous slave boson ({\it static}) mean-field study 
of the $t-J$ model.\cite{Baskaran:1987}
In fact the pseudogap size and temperature obtained in the latter
are the same as $2Js$ and $T^{MF}_{c}$ calculated from Eq.~\ref{eq10} with
$m=0$.

   In the low energy sector short-range AF correlations compete
with SC correlations to make
$T^{MF}_{c}$ or $T_{c}$ go to zero near half-filling.
At the same time, in the relatively high energy sector of order of $J$
(or in the short distance scale)
the same AF correlations induce
the local spin-singlet amplitude, which is responsible for the
normal state pseudogap in the present scenario.
When the spin-singlet (or AF spin fluctuation) aspect  
is completely neglected, 
the current approach reduces to the AF
(or SC for $T < T^{MF}_{c}$) fluctuation
scenario for the pseudogap.
O(2) SC fluctuations associated with superfluid stiffness
come into play below $T_{c}^{MF}$\cite{Corson:1999} instead of $T^{*}$,
and diverge at $T_{c}$.
The present results are robust to variations 
of $t/J=2.5-3.5$\cite{Comment40}
and small to moderate value of $t'$.

   In this paper we have considered only the local spin-singlet
amplitude induced by short-range spin correlations
and its consequences.
The complete theory of
long-range $d$-wave superconductivity is beyond the
scope of the present formulation.
How local spin-singlets (appearing at $T^{MF}_{N}$) acquire
local SC phases (at $T^{MF}_{c}$) and eventually
establish their long-range phase
coherence (at $T_{c}$) is a challenging problem
to the theory of high temperature superconductivity.

    In summary, we have shown that the
local spin-singlet amplitude with $d$-wave symmetry,
$\langle  |\Delta_{d}(0)|^{2} \rangle$, can be induced
by short-range spin correlations
even in the absence of pairing interactions.
Fluctuation theory has been formulated to make connection
between $T^{*}$, $\Delta_{pg}$
and $\langle  |\Delta_{d}(0)|^{2} \rangle$.
In the present scenario for the pseudogap,
the normal state pseudogap is caused by the induced local
spin-singlet amplitude due to short-range spin correlations,
which compete in the low energy sector with SC
correlations to make $T_{c}$ go to zero near half-filling.
Since the SC order parameter vanishes at $T_{c}$
(at $T^{MF}_{c}$ in this paper), the SC gap below $T_{c}$ is
smoothly connected to the normal state pseudogap above $T_{c}$ with
the same momentum dependence and magnitude.
Calculated $T^{*}$ falls from a high value onto the $T_{c}$ line and
closely follows $T_{N}^{MF}$.
The calculated $\Delta_{pg}$ is
in good agreement with experimental results.
We have proposed an experiment in which the present scenario
can be critically tested.
It would be interesting to see how robust
are the features found in this paper,
when the no-double-occupancy constraint is strictly imposed
on the $t-J$ model and an inhomogeneous solution is used.
%
%
%
%

   The author would like to thank Prof. A. M. Tremblay for numerous help and
discussions throughout the work. He also thanks Prof. J. L. Tallon for
sending their data, and Profs. J. W. Loram, J. L. Tallon and T. Timusk
for useful discussions.
The present work was supported by a grant from the Natural Sciences and
Engineering Research Council (NSERC) of Canada and the Fonds pour la
formation de Chercheurs et d'Aide \`a la Recherche (FCAR) of the Qu\'ebec
government.

\begin{references}
\bibitem{Comment15}
The spin-singlet correlation function (with
$d$-wave symmetry) is more general than
the $d$-wave pair correlation function
in the sense that
the spin-singlet correlations can be also induced by
short-range spin correlations even in the absence of
pairing interactions.
\bibitem{Comment30}
Exactly at half-filling the pseudogap should be replaced by 
a Mott-Hubbard gap of order $U$.
\end{references}
\end{document}